\documentclass[aps,prl,reprint,a4paper,showpacs,twocolumn]{revtex4-1}
\usepackage{amsmath,amssymb}
\usepackage{graphicx}
\usepackage{bm}

\begin{document}

\title{Typical behavior of the linear programming method for combinatorial optimization problems:\,
 From a statistical-mechanical perspective}
\date{\today}

\author{Satoshi Takabe}
\email{E-mail: s{\_}takabe@huku.c.u-tokyo.ac.jp}
\author{Koji Hukushima}
\affiliation{Graduate School of Arts and Sciences, The University of Tokyo, 3-8-1 Komaba, Meguro-ku, Tokyo 153-8902, Japan}

\begin{abstract}
Typical behavior of the linear programming problem (LP) is studied as a relaxation of the minimum vertex cover problem,
which is a type of the integer programming problem (IP).
To deal with the LP and IP by statistical mechanics, a lattice-gas model on the Erd\"os-R\'enyi random graphs
is analyzed by a replica method.
It is found that the LP optimal solution is typically equal to that of the IP
below the critical average degree $c^\ast=e$ in the thermodynamic limit.
The critical threshold for LP$=$IP is beyond a mathematical result,
 $c=1$,  and coincides with the replica-symmetry-breaking threshold of the IP. 
\end{abstract}


\maketitle

 The linear programming problem (LP) is one of the most fundamental classes of the optimization problem.
 It is a main subject in the operations research and well known as a tool for approximation of
 the integer programming problems (IP) and combinatorial optimization problems~\cite{vv}, which usually belong to a class of NP-hard.
 Some approximation rates of combinatorial optimization problems are basically determined by the integrality gap of the LP relaxation.
 In terms of approximation, two interesting issues are raised.
 One is how precisely we can approximate problems, in the worst case, with the upper or lower bound of error rates.
 Computer scientists have been addressing this worst-case issue to achieve a better approximation rate.
 The other is how precisely we can typically obtain optimal solutions by approximation algorithms.
 This issue is significant for both practical and theoretical aspects.
 The landmark work is known as the Hoffman-Kruskal's theorem on the basis of a totally unimodular matrix~\cite{hoff},
 which guarantees optimal solutions of the LP to be equivalent to those
 of the IP. 

 Let us consider randomized minimum vertex cover problems (min-VC) on Erd\"os-R\'enyi random graphs.
The min-VC 
is a type of combinatorial optimization problems and expressed as a form of the IP.
 Then,  typical performance of the LP for the min-VC is considered by
 taking an average over the random graphs 
 with the average degree $c$ fixed.
 In this case, the Hoffman-Kruskal's theorem guarantees that both the
 min-VC and its relaxed LP have similarly the same optimal values when
 $c<1$, 
where  ``similarly same'' means that they 
coincide with each other 
up to the highest order of a cardinality of variables.
 However, the previous numerical study reported that they have similarly
 the same optimal solutions below 
$c=2.62(17)$~\cite{dh}.
 This fact suggests the existence of an intrinsic correspondence between
the min-VC and its relaxed LP 
other 
than the Hoffman-Kruskal's theorem.
 Evaluating the threshold 
precisely is of significance not only in the view of typical precision of the LP but also
 to associate the typical LP performance with other properties.
 In this Letter, we shed light on the fact from a statistical-mechanical perspective.

 The statistical physics dealing with spin-glass models has been developed since 1970s~\cite{sg}.
 A mean-field approach predicts the existence of the spin-glass phase
and phase boundaries of 
spin-glass models
 such as Sherrington-Kirkpatrick model~\cite{25}.
In general, randomized 0/1 combinatorial optimization problems are
 transformed into Ising spin-glass models on sparse random graphs. 
The statistical-mechanical 
techniques 
such as a replica method and cavity  method 
have been applied to 
estimate  a typical optimal value of the
 $K$-max-SAT~\cite{22}, $K$-max-XORSAT~\cite{23},
 $\alpha$-min-VC~\cite{4,th}, and so on,  
in the level of the
 one step replica symmetry breaking ansatz~\cite{26}.
However, the LP has not been extensively studied by the
statistical-mechanical approach,  
while  
a continuous relaxation of 
randomized multidimensional knapsack problems has been
studied~\cite{inoue}. 
 Here we 
introduce a statistical-mechanical lattice-gas model which includes 
the min-VC and its relaxed LP as special cases. 
This enables us to analyze typical behavior of the LP quantitatively comparable to numerical results.
 Our analysis based on the replica method reveals that the IP and LP have similarly the same optimal solutions beyond the Hoffman-Kruskal's theorem
 from the statistical-mechanical aspect.

 The problem studied in this Letter is the min-VC. 
 It is a combinatorial optimization problem on an undirected loopless
 graph $G=(V,E)$, 
 where $V$ is a vertex set and $E\subset V^2$ is an edge set.
 We define the cardinality of 
$V$ and $E$ as $N$ and $M$, respectively.
 A vertex in the graph $G$ is labeled by $i\in\{1,\cdots,N\}$ and assigned a binary variable $x_i$.
An edge is expressed as $E_a=(i,j)\in E$ with $a\in\{1,\cdots,M\}$.
The vertex $i$ is called covered if $x_i=1$ and uncovered otherwise.
The edge connecting to at least one covered vertex is also called covered.
 The covered vertex set $V'$ is defined as a subset of $V$ which covers all the edges of $G$.
 The min-VC is a problem searching the minimum cardinality $|V'|$ of the covered vertex set.
 It is expressed in a form of the IP as 
\begin{align}
&\mathrm{Minimize}\hspace{9pt} x_c^{\mathrm{IP}}(G)=N^{-1} \bm{c}\cdot\bm{x},\nonumber\\
&\mathrm{Subject\ to}\,\, A\bm{x}\ge 1,\,\bm{x}\ge 0,\,\bm{x}\in\bm{Z}^N, \label{eq_1}
\end{align}
 where $\bm{x}=(x_1,\cdots,x_N)^{\mathrm T}$, $\bm{c}=(1,\cdots,1)^{\mathrm T}$, and
 an $M\times N$ incident matrix $A=(a_{ij})$ is defined by
 $a_{ai}=a_{aj}=1$ if $(i,j)\in E_a$ and $a_{ak}=0$ for $k\neq i,j$. 
 Our task is to find the minimum-cover ratio $x_c^{\mathrm{IP}}(G)$
 while it belongs to a class of NP-hard.
 In order to approximate it in polynomial time, the LP relaxation is executed.
 We replace 
integral conditions $\bm{x}\in\bm{Z}^N$ with 
 continuous ones $\bm{x}\in\bm{R}^N$.
 The relaxed problem may lower its optimal value $x_c^{\mathrm{LP}}(G)$ and
 contains non-integer values in a set of optimal variables
 $\{x_i^{\mathrm{opt}}\}$ in general.

 Among several algorithms for the LP such as the ellipsoid method~\cite{elip},
 we focus on the simplex method, 
which has 
a notable property of the relaxed LP for the min-VC called half-integrality~\cite{nt}.
 It has been proven that variables on extreme-point solutions of the LP take half-integer or integer, that is $0$, $1/2$, or $1$.
 The simplex method, which solves the LP to search the extreme-point solutions of a constraint polytope,
 always finds its optimal variables with integers and half-integers.
 The fact simplifies our analysis because the difference between the LP with the simplex method and the IP for the min-VC
 is whether there is a half-integer or not in the optimal variables. 
It is, however, possible to have various degenerated LP solutions with
different half-integral ratios. 
We thus introduce the half-integral ratio $p_h(G)$ as the minimum ratio
 of $1/2$ 
in the LP optimal solutions on a graph $G$.
 If $p_h(G)=0$,  
one can recognize that the LP has the same optimal solutions as the IP on the graph $G$.

 We study 
the min-VC 
defined on Erd\"os-R\'enyi 
random graphs to examine typical performance of the LP and IP.
 Edges are independently chosen from all pairs 
of vertices with
 probability $c/(N-1)$, where the average degree $c$ is average number
 of edges connecting to a vertex. 
 We define 
 a minimum-cover ratio of the LP and IP averaged over the random graphs as
 $x_c^{\mathrm{LP}}(c)$ and $x_c^{\mathrm{IP}}(c)$ in the large $N$ limit, respectively,
 which are functions of the average degree $c$. The averaged
 half-integral ratio of the LP in the limit is 
denoted by $p_h(c)$.
 If $x_c^{\mathrm{LP}}(c)=x_c^{\mathrm{IP}}(c)$ and $p_h(c)=0$, we consider that LP and IP have similarly the same optimal solutions and express it as LP$=$IP.
 The Erd\"os-R\'enyi random graph has remarkable property called a percolation~\cite{26}.
 If $c<1$, the graph contains trees with $O(\ln N)$ vertices except for cycles with finite length.
 Otherwise, the graph has a giant component with $O(N)$ vertices, which includes cycles with length $O(\ln N)$.
 The Hoffman-Kruskal's theorem guarantees that if $c<1$, the LP and IP are coincident according to our definition
 because a graph consists of trees without finite number of short cycles.
We concentrate on studying the LP for the min-VC
 on the Erd\"os-R\'enyi random graph. 

 Let us consider a statistical-mechanical model of the LP and IP for min-VCs. 
 While the half-integrality of the LP simplifies the model, we need to separate trivial ground states from solutions of the LP.
Suppose a min-VC on the simplest graph, two vertices 1 and 2 with an edge.
 Its vertex-cover constraint is $x_1+x_2\ge 1$ and the extreme-point solutions are $(x_1,x_2)=(1,0)$ {and} $(0,1)$.
 The simplex method omits other trivial ground states, which satisfy
 $x_1+x_2=1$ and $x_1x_2\neq 0$. 
 We thus add a small penalty to the model in order not to omit correct LP optimal solutions
 but to exclude the trivial ground states.

It is convenient to convert a variable $x_i$ in Eq.~(\ref{eq_1}) into a spin variable $\sigma_i$ by transformation $\sigma_i=2x_i-1$.
 The spin variable thus takes $-1$, $0$, or $1$ by 
 half-integrality.
 In order to resolve the trivial multiplicity, our Hamiltonian consists of two terms as follows
\begin{equation}
 \mathcal{H}(\underline{\sigma};s)=\sum_{i=1}^{N}\sigma_i + s\sum_{i=1}^{N}(1-\sigma_i^2), \label{eq_2}
\end{equation}
 where $\underline{\sigma}=\{\sigma_i\}$ and a penalty factor $s>0$.
 The first term is a cost function of the LP and IP.
 The second one is a penalty term to keep spin variables $\pm 1$ (or $x_i\in\mathbf{Z}$).
 As a usual lattice-gas-model expression for min-VCs~\cite{wh1}, a grand canonical partition function contains
 constraints of the min-VC and reads
\begin{equation}
 \Xi(\mu)=\sum_{\underline{\sigma}}
 e^{-\mu\mathcal{H}(\underline{\sigma};s)}\prod_{(i,j)\in E}\theta(\sigma_i+\sigma_j), \label{eq_3}
\end{equation}
 where $\theta(x)$ is a step function.
 The penalty factor $s$ must be set as 
 ground states coincide with solutions of the LP and IP in the large $\mu$ limit.
 Here we introduce a real parameter $r$ by $s=\mu^{r-1}$ {to take a limit appropriately}. 

 In order to study typical behavior, a quenched average $\overline{(\cdots)}$ according to random graphs needs to be taken.
 We use a replica trick $\overline{\ln\Xi(\mu)}=\lim_{n\rightarrow 0}(\overline{\Xi(\mu)^n}-1)/n$ because
 it is difficult to estimate an averaged grand potential $\mu^{-1}\overline{\ln\Xi(\mu)}$ directly. 
 An order parameter is introduced by the frequency distribution $c(\vec\xi)$ of a replicated vector $\vec\xi=(\xi^1,\cdots,\xi^n)$ among $N$ spins~\cite{10}.
We suppose that under the RS ansatz the order parameter is a function of $\xi=\sum_{a=1}^{n} \xi^a$ and
 $\tilde\xi=\sum_{a=1}^{n} (\xi^a)^2$. Then, the joint distribution of effective fields $h_1$ and $h_2$ is defined
 as $c(\vec\xi)=\int dP(h_1,h_2)Z^{-n}\exp(\mu h_1\xi+\mu h_2\tilde{\xi})$,
{where $Z=1+2\exp(\mu h_2)\cosh(\mu h_1)$~\cite{3st}.}
 The saddle-point equation of $P(h_1,h_2)$ reads
 \begin{align}
&P(h_1,h_2)=\sum_{k=0}^{\infty}e^{-c}\frac{c^k}{k!}\int\prod_{i=1}^{k}dP(h_1^{(i)},h_2^{(i)})\nonumber\\
&\times\delta\left(h_1+1+\sum_iu_2(h_1^{(i)},h_2^{(i)})\right)\nonumber\\
&\times\delta\left(h_2-\mu^{r-1}+\sum_i\left[u_1(h_1^{(i)},h_2^{(i)})-u_2(h_1^{(i)},h_2^{(i)})\right]\right)
, \label{eqsp1}
 \end{align}
 where $u_1=\mu^{-1}\ln\{{Z^{-1}[1+\exp(\mu h_1+\mu h_2)]}\}$ and $u_2=(2\mu)^{-1}\ln[Z^{-1}\exp(\mu h_1+\mu h_2)]$.
It is noted that the distribution still depends on the parameter $r$ of
the penalty term in the large $\mu$ limit.   
 
\begin{figure}[!tb]
\begin{center}
 \includegraphics[trim=0 0 0 0,width=1.0\linewidth]{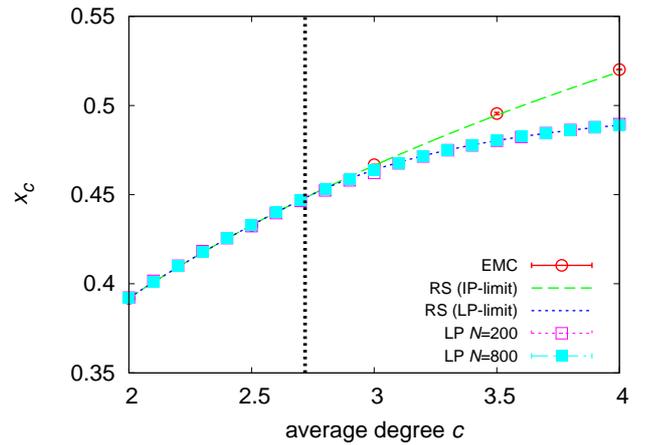}
 \caption{(Color Online). The minimum-cover ratio {in Erd\"os-R\'enyi random graphs} as a function of the average degree $c$.
{Circles} are numerical results by the {replica} exchange Monte Carlo method.
 Square marks are the data obtained by the simplex method {for the cardinality of vertices} $N=200$ (filled)
 and $N=800$ (open).
 They are averaged over $800$ random graphs.
 The {dashed and dotted} lines show the RS solutions in the IP-limit and the LP-limit, {respectively}.
 The vertical dashed line represents the critical average degree $c^\ast=e$.}
 \label{fig1}
\end{center}
\end{figure}

 We first observe an IP-limit {with} $r>1$.
 In this case, the second effective field $h_2$ diverges as $\mu\rightarrow\infty$ {and}
 the probability of $\sigma_i=0$ (or $x_i=1/2$) {vanishes} because it is proportional to $\exp(-\mu h_2)$ {for} $\mu\gg 1$.
 The saddle-point equation (\ref{eqsp1}) becomes equivalent to that of the original min-VC with binary variables~\cite{wh1}.
 This {means} that 
the IP and original min-VC 
have the same ground states of the
 model for $r>1$. 
 The RS solution becomes unstable above the critical average degree $c^\ast=e\simeq 2.71$.
 If $c<c^\ast$, we successfully estimate the minimum-cover ratio  $x_c^{\mathrm{IP}}(c)$ of the min-VC using the RS ansatz
 as shown in Fig.~\ref{fig1}.
If $c>c^\ast$, however, 
the RS estimation does not give $x_c^{\mathrm{IP}}(c)$ correctly. 

 Next, we study an LP-limit {with} $0<r<1$.
While the penalty term $\mu^{r-1}$ vanishes in the large $\mu$ limit, it
 plays an essential role in omitting configurations containing $\sigma_i=0$  (or $x_i^{\mathrm{opt}}=1/2$)
 from ground states, which are intrinsic integer solutions with $\sigma_i=\pm 1$ (or
 $x_i^{\mathrm{opt}}\in\mathbf{Z}$).
{If a ground state includes some half-integers in this limit, integral
 configurations of the model could not be the optimal one, 
 implying that the relaxed LP no longer provides the same optimal value as the original min-VC.
 Though detailed derivation will be presented in a separated paper, the self-consistent equation is referred to a simple closed form and}
 the average optimal LP value $x_c^{\mathrm{LP}}(c)$ and the average
 fraction $p_h(c)$ of 
 half-integers
 are {given} by
 \begin{equation}
 x_c^{\mathrm{LP}}(c)=1-\frac{A+B+AB}{2c},\,
 p_h(c)=\frac{(B-A)(1-A)}{c},\label{eqsp2}
 \end{equation}
 where $A$ and $B(\ge A)$ obey an equation $Ae^B=Be^A=c$.
 We find that $A=B$ below the critical average degree $c^\ast=e$.
 This leads to the LP$=$IP, 
 that is $x_c^{\mathrm{LP}}(c)=x_c^{\mathrm{IP}}(c)$ and $p_h(c)=0$ in the large $N$ limit.
{Above the critical value}, $B$ is larger than $A$ and $p_h(c)>0$.
 The optimal value $x_c^{\mathrm{LP}}(c)$ {is significantly} smaller than $x_c^{\mathrm{IP}}(c)$. 
{It is found that the RS solution is always stable against perturbation to the self-consistent equation for any value of the
 average degree $c$ in the LP limit.}

 Finally we shortly note the $3$-state limit 
with $r<0$.
 In this limit, the penalty has no influence to the effective fields and there remain trivial ground states.
 We find that the RS solution is also stable for any $c$ and the optimal value is equal to $x_c^{\mathrm{LP}}(c)$.
 The fraction of $x_i=1/2$ is, however, no longer equal to $p_h(c)$ in
 the LP-limit and also direct numerical results by the LP shown later. 

We compare the analytic evaluation with numerical results obtained by
 two algorithms, 
the Markov chain Monte Carlo method (MCMC) for the original min-VC (or
IP) and the revised simplex method for the LP. 
{As for the MCMC,} we especially use the replica exchange Monte Carlo method (EMC)~\cite{90,91} to accelerate its relaxation time.
 The EMC is {performed} on random graphs with $16\mathchar`-256$ vertices and these results are extrapolated to estimate $x_c^{\mathrm{IP}}(c)$
 in the large $N$ limit.
{In order to estimate the LP optimal value}, the LP\_solve\_5.5 solver~\cite{lpsolve} with the revised simplex method is executed.
 Both methods solve 800 random graphs for each average degree $c$ and $N$.
As shown in Fig.~\ref{fig1}, these numerical data coincide with the RS
 estimate 
in the IP-limit and LP-limit 
below the critical average degree $c^\ast=e$. 
For $c>c^*$,  the results of the LP and IP {are separated}.
 In the case of the IP, the prediction is no more appropriate when $c>e$
 because the RS solution 
is unstable.
In fact,  $x_c^{\mathrm{IP}}(c)$ 
is slightly larger than the RS estimate and grows to 1 in the large $c$ limit.
 In contrast, the numerical results of the LP agree with the analytic one even above the critical average degree.
 The LP optimal value $x_c^{\mathrm{LP}}(c)$ 
is smaller than $x_c^{\mathrm{IP}}(c)$ for $c>c^*$ and converges to $1/2$ in the large $c$ limit.

\begin{figure}[!tb]
\begin{center}
 \includegraphics[trim=0 0 0 0,width=1.0\linewidth]{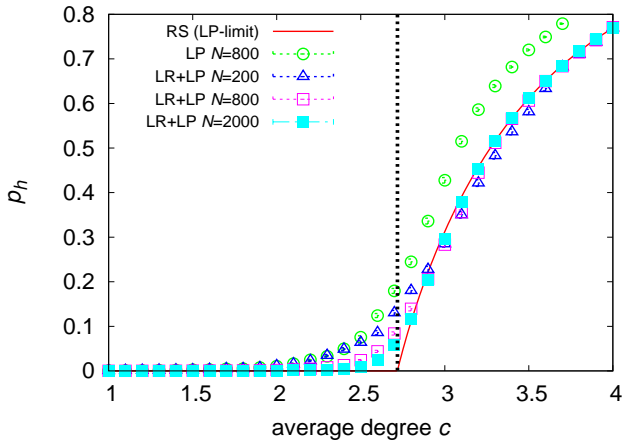}
 \caption{(Color Online). The half-integral ratio $p_h$ as a function of the average degree $c$.
{Circles} are the data obtained only by the simplex method with 
the cardinality of vertices $N=800$.
 Other open marks are numerical results by the simplex method after the LR
 with 
 $N=200$, $800$, and $2000$. 
 They are taken an average over $800$ random graphs.
 The solid line represents the RS solution in the LP-limit.
 The vertical dashed line represents the critical average degree $c^\ast=e$.}
 \label{fig2}
\end{center}
\end{figure}

\begin{figure}[!tb]
\begin{center}
 \includegraphics[trim=0 0 0 0,width=0.45\linewidth]{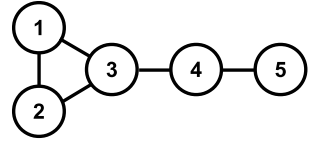}
 \caption{A graph $G_1$, which has two sets of optimal variables
 $(x_1,x_2,x_3,x_4,x_5)=(1/2,1/2,1/2,1,0)$ and $(1/2,1/2,1/2,1/2,1/2)$
 with the simplex method.} 
 \label{fig3}
\end{center}
\end{figure}

We also present the half-integral ratio $p_h(c)$ in Fig.~\ref{fig2},
where the numerical result obtained by the simplex method is
significantly larger than the RS estimate. 
It should be emphasized that $p_h(c)$ by the replica analysis is
minimized over all of the LP optimal solutions. 
{Let us consider a simple example of the graph $G_1$ in Fig.~\ref{fig3},
 which 
includes a cycle with length 3.
 LP optimal variables are found $(x_1,x_2,x_3,x_4,x_5)=(1/2,1/2,1/2,1,0)$ and $(1/2,1/2,1/2,1/2,1/2)$ by the simplex algorithm.
 Though both solutions lead to $x_c^{\mathrm{LP}}(G_1)=5/2$, their
 half-integral ratios are different, 
which might be the reason for the discrepancy found in Fig.~\ref{fig2}. 
 }
 In order to estimate $p_h(c)$ more precisely, we use the leaf removal
 algorithm (LR)~\cite{ks}, which is a graph removal algorithm for min-VCs~\cite{bg}.
 We define an original graph $G^{(0)}$ as a given graph $G$. In the $k$-th recursive step, it removes a leaf from a graph $G^{(k)}=(V^{(k)},E^{(k)})$.
 A leaf $\{v,w\}$ is a couple of vertices $v,w\in V^{(k)}$ connected by an edge $(v,w)\in E^{(k)}$ {where the} vertex $v$ is degree 1.
 Then, a new graph $G^{(k+1)}$ {includes connected components} without removed leaves and edges connecting to them.
 The LR repeats the recursive steps until there exist no leaves.
 The graph $G_{\mathrm{Fin}}$ at the end of the recursion {consists of} isolated vertices with degree 0 and a core, connected components without any leaves.
 It is proven that removed and isolated vertices by the LR are appropriately assigned covered or uncovered states.
 Thus, there is a set of optimal variables $\{x_i^{\mathrm{opt}}\}$ such that all of them are integer except for those in the core.

 To evaluate the half-integral ratio $p_h(c)$, we perform the LP on the graph $G_\mathrm{Fin}$ generated by the LR on a given graph.
{In the case of $G_1$ in Fig.~\ref{fig3}, $\{4,5\}$ is a leaf. 
The LR removes a leaf and edges $(3,4)$ and $(4,5)$.
 After the removal, the LR correctly assigns the variables $(x_4,x_5)$
 to $(1,0)$ 
and 
stops. 
 Then the simplex method solves 
the reduced graph with three vertices 1, 2 and 3.
 We finally obtain a set of optimal variables $(x_1,x_2,x_3,x_4,x_5)=(1/2,1/2,1/2,1,0)$ with the minimum half-integral ratio $3/5$.} 
 Fig.~\ref{fig2} shows that LR+LP results agree well with the analytic
 ones although 
a finite-size effect is found near $c=c^*$.
 Recall that the Hoffman-Kruskal's theorem guarantees $p_h(c)=0$ below the
 percolation threshold $c=1$ of the random graph.
{The statistical-mechanical} analysis 
leads to the larger critical average degree, which is consistent with our
 numerical results and also the previous report~\cite{dh}.

 In this Letter, we studied typical behavior of the LP, especially the
 simplex method,  using the replica method in statistical physics.
 We find the LP$=$IP phase where the optimal value of the original min-VC coincides with that of the relaxed LP
 in the order of $N$.
 Moreover, we obtain the critical average degree $c^\ast=e$ above which the LP differs from the original problem.
Interestingly, the critical degree is identical with the instability
 threshold of  the replica symmetric solution~\cite{wh1} and also the
 threshold of the LR core  percolation~\cite{bg}. 
 In terms of topology of the Erd\"os-R\'enyi {random} graph, it is notable that the half-integral ratio $p_h(c)$ is equal to the size of the LR core~\cite{bg}.
 These facts suggest that the LR core is a key to understand a replica symmetry breaking of the min-VC and
{there will exist a novel mathematical structure on random graphs}.

We demonstrate 
that the
statistical-mechanical approach enables us to predict typical
performance of the approximation algorithm. 
With the help of the
half-integrality of the min-VC, the analysis of the
 typical performance of the simplex method is reduced to study a
 statistical-mechanical model with discrete degrees of freedom.   
 It is an interesting future task to expand the proposed analysis to
 other combinatorial optimization problems without the half-integrality and 
other LP techniques such as the cutting-plane approach~\cite{sch}.
 We hope that the statistical-mechanical analysis in this Letter is of some help to understand approximation algorithms
 from a new perspective.

 This research was supported by a Grants-in-Aid for Scientific Research
 from the MEXT, 
 Japan, No. 22340109 and 25610102.

\end{document}